\documentclass[pra,amsmath,amssymb,showpacs,superscriptaddress,twocolumn]{revtex4}

\usepackage{graphicx}
\usepackage{dcolumn}
\usepackage[hypertex]{hyperref}
\usepackage{bm}
\usepackage[usenames,dvipsnames]{color}
\usepackage{amsmath}
\usepackage{amsfonts}
\usepackage{amsthm}
\usepackage{amssymb}
\usepackage{mathrsfs}
\usepackage{subfigure}
\usepackage{ulem}
\DeclareMathOperator{\sech}{sech}

\begin{document}

\title{Quantum fisher information in noninertial frames}

\author{Yao Yao}
\affiliation{Beijing Computational Science Research Center, Beijing, 100084, China}

\author{Xing Xiao}
\affiliation{Beijing Computational Science Research Center, Beijing, 100084, China}

\author{Li Ge}
\affiliation{Beijing Computational Science Research Center, Beijing, 100084, China}

\author{Xiao-guang Wang}
\email{xgwang@zimp.zju.edu.cn}
\affiliation{Zhejiang Institute of Modern Physics, Department of Physics, Zhejiang University, Hangzhou 310027, China}

\author{Chang-pu Sun}
\email{cpsun@csrc.ac.cn}
\affiliation{Beijing Computational Science Research Center, Beijing, 100084, China}

\date{\today}

\begin{abstract}
We investigate the performance of quantum fisher information under the Unruh-Hawking effect, where one of the observers (eg, Rob) is
uniformly accelerated with respect to other partners. In the context of relativistic quantum information theory, we demonstrate that
quantum fisher information, as an important measure of the information content of quantum states, has a rich and subtle physical structure
comparing with entanglement or Bell nonlocality. In this work, we mainly focus on the parameterized (and arbitrary) pure two-qubit states,
where the weight parameter $\theta$ and phase parameter $\phi$ are naturally introduced. Intriguingly, we prove that $\mathcal{F}_\theta$
keeps unchanged for both scalar and Dirac fields. Meanwhile, we observe that $\mathcal{F}_\phi$ decreases with the increase of acceleration $r$
but remains finite in the limit of infinite acceleration. More importantly, our results show that the symmetry of $\mathcal{F}_\phi$
(with respect to $\theta=\pi/4$) has been broken by the influence of Unruh effect for both cases.
\end{abstract}


\maketitle
\section{INTRODUCTION}
Quite recently, quantum fisher information (QFI) \cite{Braunstein1994,Braunstein1996} has attracted considerable attention, not only due to its own significance in
quantum estimation theory and quantum information theory, but also associated with recent rapid progress in quantum-enhanced metrology \cite{Giovanetti2004,Giovanetti2006,Giovanetti2011}.
In fact, as an important measure of information content of quantum states, QFI has already played a critical role in quantum statistical inference
through its inextricable relationship with Cram\'{e}r-Rao inequality \cite{Helstrom1976,Holevo1982}. Moreover, QFI also has various applications in other
quantum information tasks such as characterization of non-Markovianity \cite{Lu2010}, investigation of uncertainy relations \cite{Luo2000,Luo2003,Gibilisco2007a,Gibilisco2007b,Watanabe2011}
and entanglement detection \cite{Li2013}, just to name a few. Besides, since every realistic system will inevitably suffer from the decoherence induced by the coupling to its surroundings,
it is natural to explore the dynamics of QFI in all sorts of quantum noise channels \cite{Ma2011,Zhong2013}, as we do when we analyse quantum entanglement, quantum discord and Bell nonlocality.
A great deal of research has been devoted to this perspective of QFI, which is well motivated within the framework of noisy quantum metrology \cite{Escher2011,Demkowicz2011}.

On the other hand, the combination of quantum information science and relativity theory leads us to a deeper interpretation of quantum mechanics \cite{Peres2004,Crispino2008} and opens up
a new way to understand the information paradox when black holes are involved \cite{Hawking,Unruh1976,Bombelli1986,Cai2013}. In particular, what the community care about is how the Unruh-Hawking effect affects
the ``information contents" (or more especially, correlation measures) in quantum states. Therefore, diverse efforts have been made to investigate the dynamics of teleportation fidelity \cite{Alsing2003,Alsing2004},
quantum entanglement \cite{Fuentes2005,Alsing2006,Pan2008,Moradi2009,Landulfo2009,Martin2010a,Martin2010b,Martin2010c,Wang2011,Hwang2011,Xiao2011},
quantum discord \cite{Datta2009,Wang2010,Celeri2010,Tian2012,Brown2012,Doukas2013}, Bell nonlocality \cite{Friis2011,Smith2012} and some other information quantities in noninertial frames \cite{Hosler2012,Martin2012}. Intuitively, it was widely believed that the Unruh effect can only cause the degradation of quantum correlations shared between an inertial and an accelerated observer. However, Montero and Mart\'{i}n-Mart\'{i}nez have pointed
that the Unruh effect can create net quantum entanglement depending on the choice of the inertial state \cite{Montero2011}. Hence we can not merely view the Unruh effect as a \textit{usual} noise channel
since some counterintuitive and subtle phenomena will appear.

Consequently, it will be interesting to study how the relativistic effect affects the quantum fisher information or the performance of parameter estimation protocols.
However, only a few authors have attempted to address on this problem \cite{Aspachs2010,Hosler2013a,Hosler2013b,Ahmadi2013}. M. Aspachs \textit{et al.} discussed the optimal
detection of the Unruh-Hawking effect itself (eg. the acceleration parameter or Unruh temperature) and proved that Fock states can achieve the maximal QFI when considering
a scalar field in a two-dimensional Minkowski spacetime \cite{Aspachs2010}. Hosler and Kok numerically studied the NOON states based parameter estimation protocol over an Unruh channel and found
counterintuitive result that the single-rail encoding is superior to the dual rail \cite{Hosler2013a}. Furthermore, M. Ahmadi \textit{et al.} provided a framework of relativistic quantum metrology
where the relativistic effect can be viewed as a resource for quantum technologies \cite{Ahmadi2013}. In this work, we investigate the performance of quantum fisher information
for both scalar and Dirac fields in noninertial frames. Here we restrict ourselves to consider the parameterized pure two-qubit state as the initial state, where two parameters
$\theta$ and $\phi$ are introduced. Interestingly, our analytical results indicate that even in this seemingly \textit{simple} case the QFI has a rich and subtle physical structure:
(i) $\mathcal{F}_\theta$ remains invariant for both scalar and Dirac fields; (ii) for both cases $\mathcal{F}_\phi$ decreases with the increase of acceleration $r$
but approaches a finite value in the limit of infinite acceleration and more importantly, the distribution symmetry of $\mathcal{F}_\phi$
with respect to $\theta$ has been broken by the influence of Unruh effect.

The outline of the paper is as follows. In Sec. II, we review the properties of QFI and summarize the recent progress on its analytical calculation.
In Sec. III, we investigate the effect of Unruh noise on QFI of the parameterized (and arbitrary) two-qubit states in both scalar and Dirac fields.
Finally, section IV is devoted to the discussion and conclusion.
\section{Technical preliminaries of QFI}
Consider we have an N-dimensional quantum state $\rho_\lambda$ depending on an unknown parameter $\lambda$. If we intend to extract information about $\lambda$
form $\rho_\lambda$, a set of quantum measurements $\{E(\xi)\}$ should be performed. According to classical statistical theory, the quality of any measurement(s)
can be specified by a form of information called Fisher information \cite{Fisher1925,Helstrom1976,Holevo1982}
\begin{equation}
F_\lambda=\int d\xi \, p(\xi|\lambda)\left[\frac{\partial \ln p(\xi|\lambda)}{\partial\lambda}\right]^2,
\end{equation}
where $p(\xi|\lambda)=Tr[E(\xi)\rho_\lambda]$ denotes the conditional probability of acquiring the measurement result $\xi$ when the value of the parameter is $\lambda$.
Optimizing over all possible measurements, we can define the quantum fisher information \cite{Braunstein1994,Braunstein1996,Pairs2009}
\begin{equation}
\mathcal{F}_\lambda=\max_{E(\xi)}F_\lambda
\end{equation}
To move forward, we can rewrite the QFI explicitly as
\begin{equation}
\mathcal{F}_\lambda=Tr(\rho_\lambda L_{\lambda}^2)
\end{equation}
by introducing the so-called symmetric logarithmic derivative (SLD) $L_{\lambda}$ satisfying the relation
\begin{equation}
\frac{\partial\rho_\lambda}{\partial\lambda}=\frac{\rho_\lambda L_{\lambda}+ L_{\lambda} \rho_\lambda}{2}
\end{equation}
It is worth pointing out that the complete set of eigenvectors of $L_{\lambda}$ constitutes the optimal POVM
to achieve the QFI (or equivalently, the maximal classical fisher information) \cite{Pairs2009}.

Apart from the basic properties such as convexity and monotonicity (eg, nonincreasing under stochastic operations) \cite{Fujiwara2001},
the QFI is naturally related to the Bures distance or Uhlmann fidelity \cite{Braunstein1994}
\begin{equation}
d_{Bures}^2(\rho_\lambda,\rho_{\lambda+d\lambda})=\frac{1}{4}\mathcal{F}_\lambda d\lambda^2,
\end{equation}
where the Bures distance can be defined as \cite{Bures1969,Uhlmann1976,Hubner1992}
\begin{equation}
d_{Bures}(\rho,\sigma)=\sqrt{2}\left(1-Tr\sqrt{\rho^{1/2}\sigma\rho^{1/2}}\right)^{1/2},
\end{equation}
Therefore, if we have obtained the explicit form of Bures distance of the corresponding states, then the formula of QFI
can be straightforwardly derived applying this relation. Recently, we note that this strategy has already been exploited in several
situations: the single qubit \cite{Zhong2013}, single-mode Gaussian \cite{Pinel2013} and two-mode Gaussian states \cite{Ahmadi2013}.

On the other hand, based on the spectrum decomposition $\rho_\lambda=\sum_{i=1}^N p_i|\psi_i\rangle\langle\psi_i|$,
the QFI can be rephrased as \cite{Braunstein1994,Braunstein1996,Pairs2009}
\begin{equation}
\mathcal{F}_\lambda=2\sum_{m,n}^N\frac{|\langle\psi_m|\partial_\lambda\rho_\lambda|\psi_n\rangle|^2}{p_m+p_n},
\end{equation}
with the eigenvalues $p_i\geq0$ and $\sum_i^Np_i=1$. However, this expression is somewhat difficult to use especially when the dimension of $\rho_\lambda$
is very large, since the eigenvectors with respect to zero eigenvalues are all involved in the calculation. Fortunately, employing the completeness relation
\begin{equation}
\sum_{i=M+1}^N |\psi_i\rangle\langle\psi_i|=\mathbb{I}-\sum_{i=1}^M |\psi_i\rangle\langle\psi_i|,
\end{equation}
with $M$ being the dimension of the support of $\rho_\lambda$, Refs. \cite{Liu2013} and \cite{Zhang2013} provided a new expression of the QFI for general states (see also \cite{Knysh2011})
\begin{equation}
\mathcal{F}_\lambda=\sum_{i=1}^M\frac{(p'_i)^2}{p_i}+\sum_{i=1}^M4p_i\mathcal{F}_{\lambda,i}-\sum_{i\neq j}^M\frac{8p_ip_j}{p_i+p_j}|\langle\psi_i|\psi'_j\rangle|^2, \label{mixed}
\end{equation}
where $p'_i=\partial_\lambda p_i$, $|\psi'_i\rangle=|\partial_\lambda\psi_i\rangle$ and $\mathcal{F}_{\lambda,i}$ is the QFI of pure state $|\psi_i\rangle$
\begin{equation}
\mathcal{F}_{\lambda,i}=4\left(\langle\partial_\lambda\psi_i|\partial_\lambda\psi_i\rangle-|\langle\psi_i|\partial_\lambda\psi_i\rangle|^2\right). \label{pure}
\end{equation}
From this expression we can see that the QFI of a non-full-rank states is only determined by its support(eg, by the subset of $\{|\psi_i\rangle\langle\psi_i|\}$ with nonzero eigenvalues).
Furthermore, one can clearly identify that the QFI can also be divided into three parts: the first term is just the classical contribution if we regard the set of nonzero eigenvalues as a
probability distribution; the second term is a weighted average over all pure-state QFI; the last term stems from the mixture of pure states and thus decrease the
total QFI. While Eq. (\ref{mixed}) is relatively simple and has a clear physical meaning, it will be a starting point for our analysis.
\section{Quantum fisher information in noninertial frames}
In this section, we investigate how the Unruh effect affects the QFI in both scalar and Dirac fields. Since the maximally entangled states such as Bell or GHZ states
were adopted as the initial states to demonstrate the effect of Unruh noise in most previous literature, here we mainly focus on the parameterized (and arbitrary) two-qubit pure states
\begin{equation}
|\Psi_{\theta,\phi}\rangle=\cos\theta|00\rangle+e^{i\phi}\sin\theta|11\rangle, \label{initial}
\end{equation}
where the unknown arguments $\theta$ and $\phi$ are to be estimated and could be named as \textit{weight} and \textit{phase} parameters respectively.
One can easily obtain the QFI with respect to $\theta$ and $\phi$ as
\begin{equation}
\mathcal{F}_\theta=4,\quad \mathcal{F}_\phi=\sin^22\theta,
\end{equation}
which shows that the QFI of $\theta$ is irrespective of $\phi$ (and keeps constant) while the QFI of $\phi$ depends on the value of $\theta$.
Moreover, when $\theta=\pi/4$, $\mathcal{F}_\phi$ reaches its maximum indicating that the \textit{balance-weighted} state is preferable.
This observation is reminiscent of previous results about the NOON state that it has been widely exploited in quantum metrology to achieve the Heisenberg limit \cite{Giovanetti2004,Giovanetti2011}.
In addition, we also note that $\mathcal{F}_\phi$ is symmetric with respect to $\theta=\pi/4$ in the absence of Unruh effect.
\subsection{The scalar field}
To understand the Unruh effect \cite{Peres2004,Crispino2008,Unruh1976}, we assume that one observer Alice stays in a inertial frame while her partner Rob undergoes
uniform acceleration $a$, each holding a mode of a free massless scalar field in Minkowski space-time . In order to describe what Rob perceives from his perspective,
we should transform from the Minkowski coordinates $\{t,z\}$ to
the Rindler coordinates $\{\tau,\xi\}$
\begin{equation}
\begin{split}
t=a^{-1}e^{a\xi}\sinh a\tau, \  z=a^{-1}e^{a\xi}\cosh a\tau, \quad\mathbb{I} \\
t=-a^{-1}e^{a\xi}\sinh a\tau, \  z=-a^{-1}e^{a\xi}\cosh a\tau,\quad\mathbb{II}
\end{split}
\end{equation}
which defines the right (Region $\mathbb{I}$) and left (Region $\mathbb{II}$) Rindler wedges. Usually we refer to the accelerating observer in Region $\mathbb{I}$
and $\mathbb{II}$ as Rob and anti-Rob respectively. According to the Bogoliubov transformation \cite{Wald1994,Walls1994}, the Minkowski vacuum state and single excitation state
can be expressed in terms of Rindler modes
\begin{align}
|0_k\rangle^{\mathcal{M}}&=\frac{1}{\cosh r}\sum_{n=0}^\infty\tanh^n r |n_k\rangle_I|n_k\rangle_{II}, \nonumber\\
|1_k\rangle^{\mathcal{M}}&=\frac{1}{\cosh^2 r}\sum_{n=0}^\infty\tanh^n r \sqrt{n+1}|(n+1)_k\rangle_I|n_k\rangle_{II}.
\label{transformation}
\end{align}
where $|n_k\rangle_I$ and $|n_k\rangle_{II}$ denote the mode decomposition in region $\mathbb{I}$ and $\mathbb{II}$ respectively, and
$\cosh r=(1-e^{-2\pi\Omega})^{-1/2}$ with $\Omega=|k|c/a$. Since Rob is causally disconnected from Region $\mathbb{II}$ (eg, anti-Rob is physically inaccessible), we must
trace over the state of Region $\mathbb{II}$. Therefore, one can verify that an entangled pure state seen by Alice in inertial frame will be recognized as a mixed
state by the accelerated observer Rob. Therefore, in the language of quantum formation theory, the Unruh effect
can be viewed as a noise channel which consists of two steps: (i) mapping the state in Minkowski space to Rindler modes according to
the transformation (\ref{transformation}) and then (ii) tracing over the causally disconnected Region $\mathbb{II}$.

Assume that Alice and Rob initially share the two-qubit pure state Eq. (\ref{initial}) from an inertial perspective, and then Rob
is uniformly accelerated with $a$. Due to the action of the Unruh channel, we will arrive at a mixed state
\begin{equation}
\rho_{AR}=\frac{1}{\cosh^2 r}\sum_{n=0}^\infty\tanh^{2n} r\rho_n,
\end{equation}
where
\begin{align}
\rho_n=&\cos^2\theta|0,n\rangle\langle0,n|+\frac{(n+1)\sin^2\theta}{\cosh^2 r}|1,n+1\rangle\langle1,n+1| \nonumber\\
&+\frac{\sqrt{n+1}\sin\theta\cos\theta}{\cosh r}\left(e^{-i\phi}|0,n\rangle\langle1,n+1|+H.c.\right),
\end{align}
with $|n,m\rangle=|n_A\rangle^{\mathcal{M}}|m_R\rangle_I$. Further, we notice that the $2\times2$ block composed by the basis
$\{|0,n\rangle,|1,n+1\rangle\}$ can be effectively regarded as a \textit{pure qubit} since one-qubit pure states $\rho=\frac{1}{2}(I+\vec{n}\cdot\vec{\sigma})$ have the form
\begin{align}
\rho=&\frac{1}{2}
\left(\begin{array}{cc}
1+n_3 & n_1-in_2 \\
n_1+in_2 & 1-n_3
\end{array}\right)\nonumber\\
=&\frac{1}{2}
\left(\begin{array}{cc}
1+n_3 & e^{-i\vartheta}\sqrt{1-n_3^2} \\
e^{i\vartheta}\sqrt{1-n_3^2} & 1-n_3
\end{array}\right)
\end{align}
where $\vec{n}=\{n_1,n_2,n_3\}$ is the Bloch vector and $\vartheta=\arctan n_2/n_1$.
Correspondingly, $\rho_{AR}$ can be rewritten as (note that $\rho_n$ is unnormalized)
\begin{align}
\rho_{AR}&=\bigoplus_{n=0}^\infty P_n |\Phi_n\rangle\langle\Phi_n| \nonumber\\
&=\frac{1}{\cosh^2 r}\sum_{n=0}^\infty\tanh^{2n} r \,\Theta_n |\Phi_n\rangle\langle\Phi_n|,
\end{align}
where
\begin{gather}
P_n=\frac{\tanh^{2n}r}{\cosh^2 r}\Theta_n,\ \Theta_n=\cos^2\theta+\frac{(n+1)\sin^2\theta}{\cosh^2 r}, \\
|\Phi_n\rangle=\{\frac{\cos\theta}{\sqrt{\Theta_n}},\ e^{i\phi}\frac{\sqrt{n+1}\sin\theta}{\cosh r\sqrt{\Theta_n}}\}^T
\end{gather}
Therefore, the overall effect of the Unruh channel is mapping the pure state $|\Psi_{\theta,\phi}\rangle$ into a \textit{mixture} of
pure states in distinct blocks.

First, we consider the QFI associated with $\theta$. Keeping in mind that $\Theta_n$ also contains information about $\theta$,
we obtain the classical part of $\mathcal{F}_\theta$
\begin{equation}
F_C=\sum_{n=0}^\infty\frac{1}{P_n}\left(\frac{\partial P_n}{\partial\theta}\right)^2=\sum_n\frac{\tanh^{2n}r}{\cosh^2 r}\frac{(\partial_\theta\Theta_n)^2}{\Theta_n},
\end{equation}
Meanwhile, utilizing the identities
\begin{align}
\partial_\theta\Theta_n=&\sin2\theta\left(\frac{n+1}{\cosh^2 r}-1\right), \nonumber\\
=&2\sqrt{\Theta_n}\partial_\theta\left(\sqrt{\Theta_n}\right), \label{identity}
\end{align}
and the formula Eq. (\ref{pure}), we can get the quantum part of $\mathcal{F}_\theta$
\begin{align}
F_Q=&\sum_{n=0}^\infty4P_n\mathcal{F}_\theta\left(|\Phi_n\rangle\langle\Phi_n|\right) \nonumber\\
=&\sum_{n=0}^\infty\frac{\tanh^{2n}r}{\cosh^2 r}\left[4\Lambda_n-\frac{(\partial_\theta\Theta_n)^2}{\Theta_n}\right],
\end{align}
where
\begin{equation}
\Lambda_n=\sin^2\theta+\frac{(n+1)\cos^2\theta}{\cosh^2 r},
\end{equation}
It is worth emphasizing that the last term of Eq. (\ref{mixed}) does not contribute to $\mathcal{F}_\theta$ because
$|\Phi_i\rangle$ and $|\Phi_j\rangle$ with $i\neq j$ locate on different subspaces. Indeed, one can find that
$\langle\Phi_i|\partial_\theta\Phi_i\rangle$ is also equal to zero by using the identities Eqs. (\ref{identity}) again, that is
\begin{equation}
\langle\Phi_i|\partial_\theta\Phi_j\rangle=0,\ \forall \  i,j
\end{equation}
All together, $\mathcal{F}_\theta$ is given by
\begin{equation}
\mathcal{F}_\theta=F_C+F_Q=\frac{1}{\cosh^2 r}\sum_{n=0}^\infty4\tanh^{2n} r\Lambda_n,
\end{equation}
Moreover, we note that the following formulas hold
\begin{align}
\sum_{n=0}^\infty\tanh^{2n}r=&\cosh^2 r, \nonumber\\
\sum_{n=0}^\infty(n+1)\tanh^{2n}r=&\cosh^4 r,
\end{align}
Intriguingly, we finally arrive at the conclusion that $\mathcal{F}_\theta$ keeps \textit{invariant}, that is, $\mathcal{F}_\theta=4$ independent of
the acceleration parameter $r$. This result implies that the QFI about $\theta$ is unaffected under the effect of the Unruh channel.

On the other hand, $P_n$ or $\Theta_n$ do not rely on the parameter $\phi$ and then the classical part of $\mathcal{F}_\phi$ is just zero.
Therefore, $\mathcal{F}_\phi$ can only stem from the average QFI of the pure states $|\Phi_n\rangle\langle\Phi_n|$
\begin{align}
\mathcal{F}_\phi=&\sum_{n=0}^\infty4P_n\mathcal{F}_\phi\left(|\Phi_n\rangle\langle\Phi_n|\right) \nonumber\\
=&\frac{\sin^2 2\theta}{\cosh^4 r}\sum_{n=0}^\infty\frac{(n+1)\tanh^{2n} r}{\Theta_n}, \label{2F1}
\end{align}
Furthermore, we can obtain the analytical expression of $\mathcal{F}_\phi$ by introducing the hypergeometric function \cite{Abramowitz} (for simplicity, we put the formula in
the Appendix).

\begin{figure}[htbp]
\begin{center}
\includegraphics[width=0.35\textwidth ]{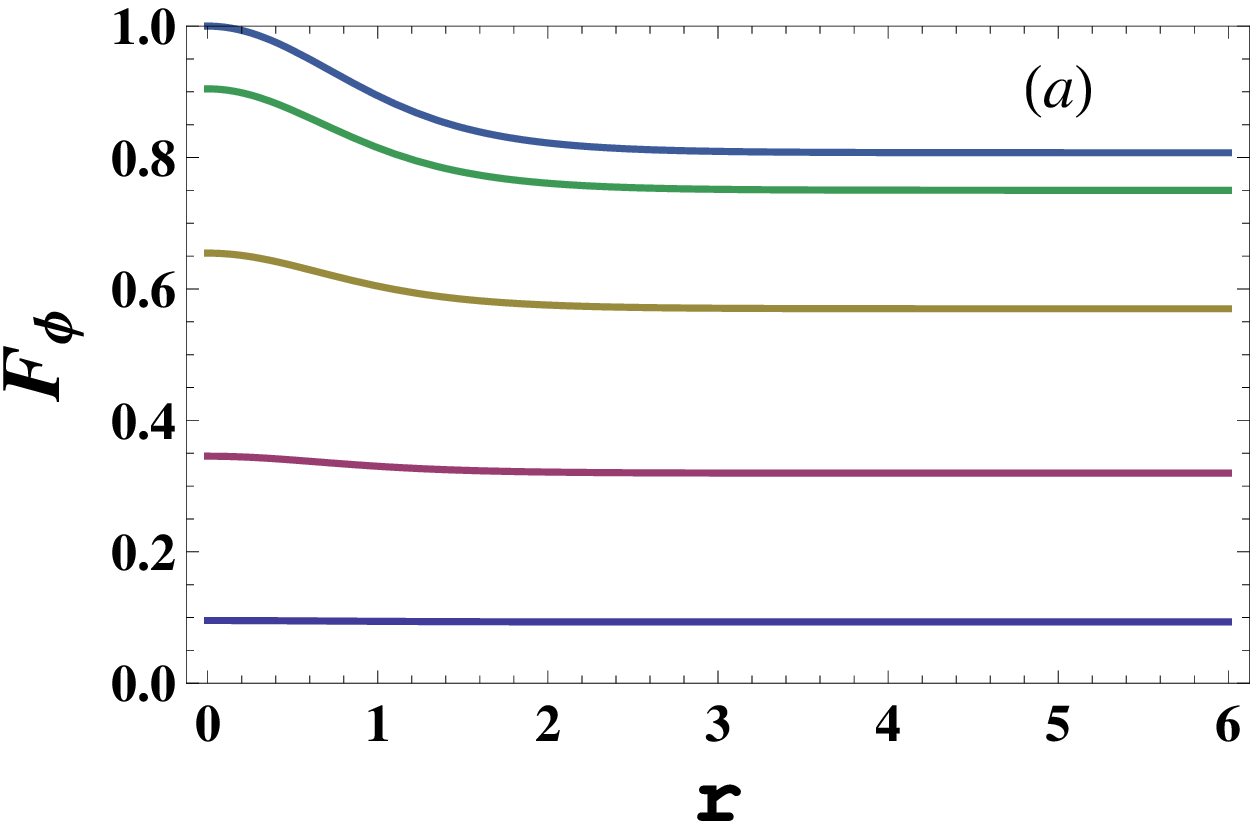}\\
\includegraphics[width=0.35\textwidth ]{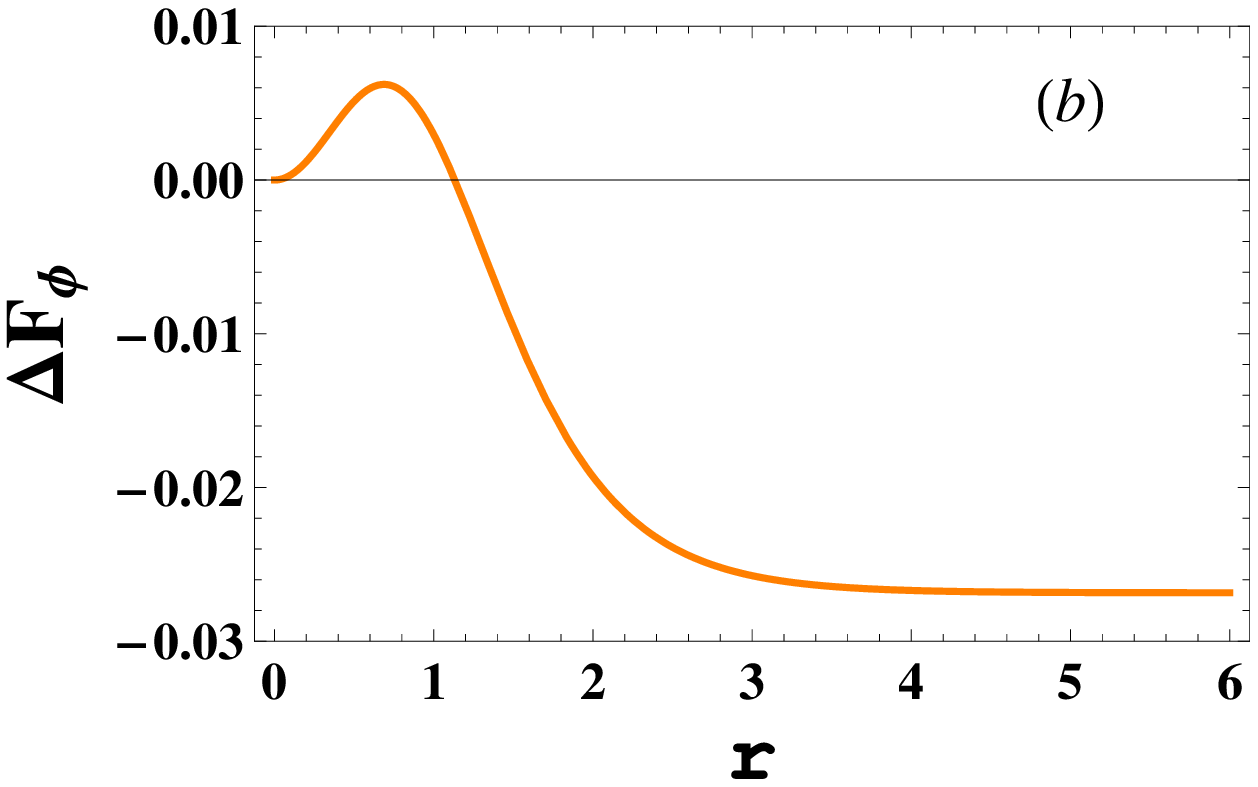}
\end{center}
\caption{(Color online) (a) The QFI $\mathcal{F}_\phi$ as a function of the acceleration parameter $r$ where $\theta=\pi/20, \pi/10,
3\pi/20, \pi/5, \pi/4$ from bottom to top; (b) $\Delta\mathcal{F}_\phi=\mathcal{F}_\phi(\theta=\pi/3)-\mathcal{F}_\phi(\theta=\pi/6)$
as a function of the acceleration $r$. Note that $\pi/6$ and $\pi/3$ are symmetric with respect to $\pi/4$.
}\label{fisher1}
\end{figure}

One can check that the limitation $\lim_{r\rightarrow0}\mathcal{F}_\phi=\sin^22\theta$ consistent with the initial value and indeed there
are some other points that deserve our attention: (i) we have plotted the $r$-dependence of $\mathcal{F}_\phi$ for a series of values of $\theta$ in Fig. \ref{fisher1}(a) and found that
as the acceleration $r$ increases, $\mathcal{F}_\phi$ gradually decreases and converges to \textit{non-zero} value in the limit of infinite acceleration, which is in sharp contrast with entanglement
\cite{Fuentes2005}; (ii) we also observe that $\mathcal{F}_\phi$ is no longer a monotonically increasing function of $\theta$ in the interval $[0,\pi/4]$ and does not
achieve the maximum at $\theta=\pi/4$ for certain fixed value of $r$. More remarkably, the symmetry of the function $\mathcal{F}_\phi$ with respect to $\theta=\pi/4$ has been broken as we can see from Fig. \ref{fisher1}(b). 
For vanishing acceleration, $\Delta\mathcal{F}_\phi=\mathcal{F}_\phi(\theta=\frac{\pi}{3})-\mathcal{F}_\phi(\theta=\frac{\pi}{6})=0$ as we expect. However, for small values of $r$, $\Delta\mathcal{F}_\phi>0$ and
when $r\rightarrow\infty$, $\Delta\mathcal{F}_\phi<0$ and also approaches a finite value.

\subsection{The Dirac field}
Now we turn to discuss the QFI for a free Dirac field. A similar analysis shows that the Minkowski vacuum and one-excitation state in terms of Rindler states
is given by \cite{Alsing2006}
\begin{gather}
|0_k\rangle^{\mathcal{M}}=\cos r |0_k\rangle_I|0_{-k}\rangle_{II}+\sin r |1_k\rangle_I|1_{-k}\rangle_{II}, \nonumber\\
|1_k\rangle^{\mathcal{M}}=|1_k\rangle_I|0_{-k}\rangle_{II},
\label{transformation1}
\end{gather}
where $\cos r=(1+e^{-2\pi\omega c/a})^{-1/2}$ (note that $\omega=|k|$) and the range of the acceleration parameter is $r\in[0,\pi/4)$ corresponding to $a\in[0,+\infty)$.
The distinction between the transformations Eqs. (\ref{transformation}) and (\ref{transformation1}) pertaining respectively
to scalar and Dirac fields is physically induced by the differences between the Bose-Einstein and Fermi-Dirac statistics \cite{Alsing2006}.

Through the Unruh channel for the Dirac case, the reduced density matrix between Alice and Rob takes the form
\begin{align}
\rho_{AR}=&\cos^2r\cos^2\theta|00\rangle\langle00|+\sin^2r\cos^2\theta|01\rangle\langle01| \nonumber\\
&+\frac{1}{2}\cos r\sin2\theta(e^{-i\phi}|00\rangle\langle11|+H.c.) \nonumber\\
&+\sin^2\theta|11\rangle\langle11|,
\end{align}
With the help of the formula Eq. (\ref{mixed}), we only need the nonzero eigenvalues
\begin{equation}
\lambda_1=1-\sin^2r\cos^2\theta,\ \lambda_2=\sin^2r\cos^2\theta,
\end{equation}
and the corresponding (normalized) eigenvectors
\begin{gather}
|\Phi_1\rangle=\frac{1}{\sqrt{1+\cos^2r\cot^2\theta}}\left\{e^{-i\phi}\cos r\cot\theta,0,0,1\right\},\nonumber\\
|\Phi_2\rangle=\left\{0,1,0,0\right\},
\end{gather}
Now we have the full information to calculate the QFI. First, the classical contribution of $\mathcal{F}_\theta$ is
\begin{align}
F_C=&\frac{(\partial_\theta\lambda_1)^2}{\lambda_1}+\frac{(\partial_\theta\lambda_2)^2}{\lambda_1}=\frac{1}{\lambda_1\lambda_2}(\partial_\theta\lambda_1)^2 \nonumber\\
=&\frac{4\sin^2r\sin^2\theta}{1-\sin^2r\cos^2\theta},
\end{align}
Meanwhile, we notice that $|\Phi_2\rangle$ does not contain any information about $\theta$ and $\phi$. Hence we can obtain the quantum part of $\mathcal{F}_\theta$
\begin{align}
F_Q=4\lambda_1\langle\partial_\theta\Phi_1|\partial_\theta\Phi_1\rangle=\frac{4\cos^2r}{1-\sin^2r\cos^2\theta},
\end{align}
where we resort to the facts that $\partial_\theta(\cot\theta)=-\csc^2\theta$ and $\langle\Phi_1|\partial_\theta\Phi_1\rangle=0$.
Therefore, the QFI with respect to $\theta$ is given by
\begin{align}
\mathcal{F}_\theta=F_C+F_Q=4,
\end{align}
Interestingly, we have shown that $\mathcal{F}_\theta$ also remains unchanged which is the same as the scalar case.
It is worth stressing that this is a highly nontrivial result, since for an arbitrary single-qubit state
W. Zhong \textit{et.al.} found that only phase-damping channel would cause no disturbance to $\mathcal{F}_\theta$ \cite{Zhong2013}.

\begin{figure}[htbp]
\begin{center}
\includegraphics[width=.35\textwidth]{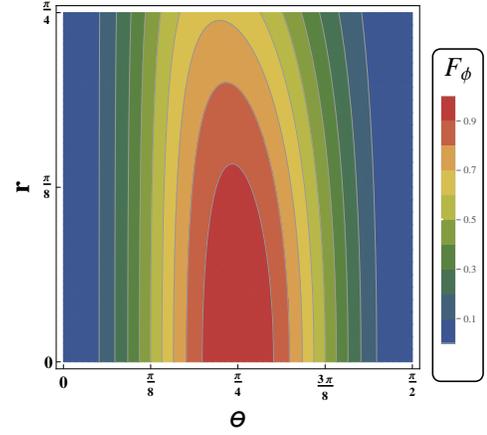} {}
\end{center}
\caption{(Color online) The contour plot of $\mathcal{F}_\phi$ as a function of the acceleration parameter $r$ and $\theta$.}
\label{fisher2}
\end{figure}

Correspondingly, the QFI $\mathcal{F}_\phi$ boils down to
\begin{align}
\mathcal{F}_\phi=&4\lambda_1\left(\langle\partial_\phi\Phi_1|\partial_\phi\Phi_1\rangle-|\langle\Phi_1|\partial_\phi\Phi_1\rangle|^2\right) \nonumber\\
=&\frac{\cos^2r\sin^22\theta}{1-\sin^2r\cos^2\theta},
\end{align}
Similarly, one can easily find that $\lim_{r\rightarrow0}\mathcal{F}_\phi=\sin^22\theta$. Interestingly, the symmetry with respect to $\theta=\pi/4$
is also broken due to the Unruh effect (see Fig. \ref{fisher2}), the same as the situation of the scalar field.

However, one can prove that
$\mathcal{F}_\phi$ can \textit{not} be amplified along with increasing the acceleration. In fact, the first order derivative of $\mathcal{F}_\phi$ is given by
\begin{align}
\frac{\partial\mathcal{F}_\phi}{\partial r}=-\frac{4\sin2r\sin^4\theta\cos^2\theta}{(1-\sin^2r\cos^2\theta)^2}\leq0,
\end{align}
And yet, $\mathcal{F}_\phi$ does not reduce to zero but remains finite in the infinite acceleration limit, that is
\begin{align}
\lim_{r\rightarrow\pi/4}\mathcal{F}_\phi=\frac{1-\cos4\theta}{3-\cos2\theta},
\end{align}

Finally, we also investigate how the Unruh effect influences the distribution of the QFI over subsystems \cite{Lu2012}. It is straightforward to
show that
\begin{gather}
\mathcal{F}_\theta^A=4,\ \mathcal{F}_\theta^R=\frac{4\cos^2r\sin^2\theta}{1-\cos^2r\cos^2\theta},\nonumber\\
\mathcal{F}_\phi^A=\mathcal{F}_\phi^R=0.
\end{gather}
where $\mathcal{F}_\xi^X$ denotes the QFI with respect to parameter $\xi\in\{\theta,\phi\}$ in subsystem $X\in\{A,R\}$.
One can easily check that the reduced states $\rho^A$ and $\rho^R$ contain no information about $\phi$ and hence $\mathcal{F}_\phi\geq\mathcal{F}_\phi^A+\mathcal{F}_\phi^R$
irrespective of the values of $r$ and $\theta$. Since $\lim_{r\rightarrow0}\mathcal{F}_\theta^R=\mathcal{F}_\theta=4$ and $\mathcal{F}_\theta^R\geq0$, actually $\mathcal{F}_\theta$ is always smaller than the sum of QFI
of the two subsystems. This is to say, the QFI can be either subadditive or superadditive under the Unruh effect depending on the specific parameter that we concern.

\section{DISCUSSION AND CONCLUSION}
As an important measure of the information content of quantum states, quantum fisher information plays an essential role in both statistical theory
and quantum metrology. However, unlike other correlation measures such as quantum entanglement and discord, the QFI usually reveals intricate and subtle behavior
since the QFI characterizes the intrinsic sensitivity of the system being discussed with respect to the change of certain specific parameters.
In this paper, we have investigated the performance of QFI under the Unruh-Hawking effect in the context of relativistic quantum information theory.
Here we mainly focus on the parameterized (and arbitrary) pure two-qubit states where two corresponding parameters are naturally introduced: the weight parameter $\theta$
and phase parameter $\phi$. We assume that Alice and Rob each share one mode of the two-qubit state, and then Rob is uniformly accelerated. Meanwhile, from the perspective of
quantum information theory, the Unruh effect can also be viewed as a particular quantum operation or quantum noise channel \cite{Aspachs2010,Bradler2009}.

To complete our analysis, both bosonic and fermionic field are detailedly considered. Our analytical results indicates that the QFI has a rich and subtle physical structure
in both cases. For the scalar field, $\mathcal{F}_\theta$ keeps constant irrespective of the acceleration parameter $r$, while $\mathcal{F}_\phi$ decreases gradually with
increasing acceleration but remains finite in the $r\rightarrow\infty$ limit. For the Dirac field, we have demonstrated that $\mathcal{F}_\theta$ still
remains unchanged, similarly to the bosonic case. Meanwhile, we have proved that $\mathcal{F}_\phi$ can never be amplified for arbitrary values
of $r$ and $\theta$. Furthermore, we find that the symmetry of $\mathcal{F}_\phi$ with respect to $\theta=\pi/4$ is broken by the Unruh effect in both cases.
Finally, we have also checked how the Unruh effect affects the distribution of the QFI over subsystems of a Dirac field. It is demonstrated that
$\mathcal{F}_\phi$ is obviously superadditive but $\mathcal{F}_\theta$ is always subadditive independent of the acceleration $r$ and $\theta$ itself \cite{Lu2012}.

In view of these findings, there are some further problems to be addressed: (i) the relativistic effect on quantum metrology and other quantum technologies is still not so clear.
Indeed, for specific setting-ups, one should clarify whether the relativistic effect could be a resource or an obstacle for corresponding quantum tasks \cite{Ahmadi2013};
(ii) The distribution property of QFI in tripartite (eg, GHZ states) or multipartite systems would be very interesting to investigate, especially when certain kinds of quantum channels
(or quantum noises) are introduced. Moreover, when finalizing the work, we are aware that our analytical results can be regarded as a complement to the recent numerical work
by Hosler and Kok \cite{Hosler2013a} (see Fig. 2 and 3 in their paper).

\begin{acknowledgments}
This work was supported by the National Natural Science Foundation of China (Grant Nos. 11121403, 10935010, 11074261 and 11247006) and
the National 973 program (Grant No. 2012CB922104 and No. 2014CB921402).
\end{acknowledgments}
%%%%%%%%%%%%%%%%%%%%%%%%%%%%%%%%%%%%%%%%%%%%%%%%%%%%%%%%%%%%%%%%%%%%%%%%%%%%%%%%%%%%%%%%%%%%%
\appendix
\section{The analytical expression of $\mathcal{F}_\phi$}
Here, we present the analytical expression of the summation of the series Eq. (\ref{2F1}).
\begin{widetext}
\begin{equation}
\mathcal{F}_\phi=\frac{\sech^4r\sin^22\theta\left[{}_{2}F_1 (a_1,b_1;c_1;z_1)(\cos^2\theta+2\sech^2r\sin^2\theta)+{}_{2}F_1 (a_2,b_2;c_2;z_2)(\cos^2\theta+\sech^2r\sin^2\theta)\tanh^2r\right]}
{\cos^4\theta + 3 \sech^2r \sin^2\theta\cos^2\theta  + 2 \sech^4r \sin^4\theta},
\end{equation}
where
\begin{align}
a_1=1,\ b_1=1+\cosh^2r\cot^2\theta,\ c_1=2+\cosh^2r\cot^2\theta,\ z_1=\tanh^2r; \nonumber\\
a_2=2,\ b_2=2+\cosh^2r\cot^2\theta,\ c_2=3+\cosh^2r\cot^2\theta,\ z_2=\tanh^2r,
\end{align}
\end{widetext}
The hypergeometric function ${}_{2}F_1 (a,b;c;z)$ is defined for $|z|<1$ by the power series
\begin{equation}
{}_{2}F_1 (a,b;c;z)=\sum_{n=0}^\infty\frac{(a)_n(b)_n}{(c)_n}\frac{z^n}{n!},
\end{equation}
where $(q)_n$ is Pochhammer symbol
\begin{align}
(q)_n=\left\{\begin{array}{cc}
1  & \mbox{ if } \, n=0 \\
q(q+1)\cdots(q+n-1) & \mbox{ if } \, n>0
\end{array}\right.
\end{align}
For more details about the hypergeometric function, we refer the readers to Ref. \cite{Abramowitz}.

%%%%%%%%%%%%%%%%%%%%%%%%%%%%%%%%%%%%%%%%%%%%%%%%%%%%%%%%%%%%%%%%%%%%%%%%%%%%%%%%%%%%%%%%%%%%%

\end{document}